\documentclass[aps,pra,twocolumn,superscriptaddress]{revtex4}
\usepackage{filecontents}
\usepackage{amssymb}
\usepackage{graphicx}
\usepackage{amsmath}
\usepackage{xcolor}
\newcommand\beq{\begin{equation}}
\newcommand\eeq{\end{equation}}
\newcommand\ket[1]{\left|#1\right\rangle}
\newcommand\de{{\delta}}
\newcommand\ka{{\kappa}}
\newcommand\la{\lambda}
\newcommand\Si{{\Sigma}}
\newcommand\vs{{\it vs.}\ }
\newcommand\eps{\epsilon}
\newcommand\bea{\begin{eqnarray}}
\newcommand\eea{\end{eqnarray}}

\newcommand{\ba}{\begin{align}}
\newcommand{\ea}{\end{align}}

\newcommand\epsbar{{\eps_0}}

\newcommand\hdd{H^\text{DD}}
\newcommand\hssh{H_\text{SSH}}
\newcommand\hinf{H_\infty}
\newcommand\hsshi{H_\text{SSH,$\infty$}}
\newcommand{\rc}{{r_{\text{C}}}}
\newcommand{\tc}{{t_{\text{C}}}}
\newcommand{\tl}{{t_{\text{L}}}}
\newcommand{\Sii}{\Si_\infty}
\newcommand{\Sisshi}{\Si_{\text{SSH},\infty}}
\newcommand{\Gsshi}{G_{\text{SSH},\infty}}
\newcommand{\GSsshi}{G^{\text{S}}_{\text{SSH},\infty}}

\newcommand\ignorethis[1]{{}}

\begin{document}

\title{Detecting topological edge states with the dynamics of a qubit}

\author{Meri Zaimi}
\email{meri.zaimi@umontreal.ca}
\affiliation{Centre de Recherches Math{\'e}matiques, Universit{\'e} de Montr{\'e}al, C.P. 6128, succursale Centre-ville, Montr{\'e}al, QC H3C 3J7, Canada}
\author{Christian Boudreault}
\email{Christian.Boudreault@cmrsj-rmcsj.ca}
\affiliation{D\'{e}partement des sciences de la nature, Coll\`{e}ge militaire royal de Saint-Jean
15 Jacques-Cartier Nord, Saint-Jean-sur-Richelieu, QC Canada, J3B 8R8}
\author{Nou\'edyn Baspin}
\email{nouedyn.baspin@mail.mcgill.ca}
\affiliation{Department of Physics, McGill University, Montr\'{e}al, QC, Canada, H3A 2T8}
\author{Nicolas Delnour}
\email{nicolas.delnour@umontreal.ca}
\affiliation{D{\'e}partement de physique, Universit{\'e} de Montr{\'e}al, Complexe des Sciences,  C.P. 6128, succursale Centre-ville, Montr{\'e}al, QC, Canada, H3C 3J7}
\author{Hichem Eleuch}
\email{heleuch@fulbrightmail.org}
\affiliation{Department of Applied Physics and Astronomy, University of Sharjah, Sharjah, UAE}
\author{Richard MacKenzie}
\email{richard.mackenzie@umontreal.ca}
\affiliation{D{\'e}partement de physique, Universit{\'e} de Montr{\'e}al, Complexe des Sciences,  C.P. 6128, succursale Centre-ville, Montr{\'e}al, QC, Canada, H3C 3J7}
\author{Michael Hilke}
\email{hilke@physics.mcgill.ca}
\affiliation{Department of Physics, McGill University, Montr\'{e}al, QC, Canada, H3A 2T8}

\begin{abstract}
We consider the Su-Schrieffer-Heeger (SSH) chain, which has 0, 1, or 2 topological edge states depending on the ratio of the hopping parameters and the parity of the chain length. We couple a qubit to one edge of the SSH chain and a semi-infinite undimerized chain to the other, and evaluate the dynamics of the qubit. By evaluating the decoherence rate of the qubit we can probe the edge states of the SSH chain. The rate shows strong even-odd oscillations as a function of site number, reflecting the presence or absence of edge states. Hence, the qubit acts as an efficient detector of the topological edge states of the SSH model. This can be generalized to other topological systems.
\end{abstract}

\maketitle

\section{Introduction}
Qubits are the building blocks of any quantum information processing device. Two of the most challenging problems for quantum computing and other applications are decoherence due to the interaction with environment and perturbations due to manufacturing imperfections \cite{Steane_1998,Nielsen2000,PhysRevLett.122.014103}. These effects limit the effective performance of quantum devices, such as the speed of an eventual quantum computer. Thus, evaluating the decoherence rate for the qubit or for an ensemble of coupled qubits is of great importance.

In previous work \cite{eleuch2017probing}, the decay rate of a qubit coupled to another system with or without disorder was studied. The main objective was to investigate under which circumstances the interaction of a qubit with its surroundings can be designed to improve the qubit's performance in a quantum device by increasing the decoherence time. It was shown that the decoherence rate of the qubit is related to transport properties of the coupled system. Furthermore, it was proven that disorder lowers the decoherence rate on average. This suggests potential applications to increase the performance of qubits in a quantum device by adding impurities to the system.

In this work, a similar composite system is studied from a different perspective. Rather than viewing the qubit as the system of interest and tailoring the system with which it interacts to improve the qubit's decoherence, the qubit is viewed as a measuring device capable of determining properties of its environment. In particular, we explore the dynamics of a qubit attached to a Su-Schrieffer-Heeger (SSH)  chain which is then attached to a third system modeling the environment; the third system is a standard tight-binding hopping Hamiltonian (without dimerization). 

The SSH model, described in detail below, is one of the simplest systems exhibiting interesting topology such as solitons and, of interest here, edge states \cite{heeger1988solitons,atala2013direct,wang2013topological,lohse2016thouless,nakajima2016topological,leder2016real}.
In spite of the model's inherent simplicity, it manages to capture many interesting and important physical effects in topological systems. The model has also been extended to study topological insulators of higher dimensions \cite{xie2019topological}.
Normally, almost-zero-energy edge states have exponentially localized
wavefunctions at the edges. These states are a particular type of topological edge states.
Topological edge states have captured the interest of researchers in in several fields of physics due to their diverse surprising proprieties. To name but a few examples, they can enhance the sound intensity at phononic crystal interfaces \cite{nature2015}, allow a robust one-way propagation \cite{Nature2016} or protect light transport in nanophotonics systems \cite{Nature2019}. For further applications and references, see \cite{EE1,EE2}.

We evaluate the decoherence rate of the qubit and how it depends on the properties of the SSH system in order to probe the edge states of the SSH chain. As we will see, it is strongly affected by such states at the qubit end of the coupled system.

In the next section, we review the isolated SSH model, mainly to establish notation but also to highlight the conditions for the existence of edge states and their properties. In Section III, we explore the double dot coupled to an SSH chain which is itself coupled to semi-infinite chain. An expression for the decoherence rate is derived using a semi-analytic approximation which is in excellent agreement with numerical simulations of the same system. We will see a strong effect of edge states on the decoherence rate. We conclude with a discussion of our results and avenues for future work in Section IV.

\section{Su-Schrieffer-Heeger model}
The SSH model \cite{PhysRevLett.42.1698} is a one-dimensional tight-binding model with alternating coupling strengths due to the Peierls instability \cite{peierls1,peierls2},
leading to a parity effect in the chain length. The Hamiltonian for a chain of $N$ sites is
\beq
\hssh=\left(\begin{matrix}
0 & t_1 \\
t_1 & 0 & t_2 &\phantom{\ddots}\\
& t_2 & 0 & t_1 &\phantom{\ddots} \\
& & t_1 & 0 & \ddots & \phantom{\ddots} \\
\phantom{\ddots}&\phantom{\ddots} &\phantom{\ddots} & \ddots & \ddots & t \\
& & & & t & 0 \\
\end{matrix}\right),
\label{eq-H_SSH}
\eeq
where $t=t_1$ or $t_2$ for $N$ even or odd, respectively. We will assume $t_1,t_2>0$ for simplicity.

\subsection{Overview of the SSH model solutions}

\begin{figure}[h!]
	\centering
	\hspace*{0cm}\includegraphics[width=0.47\textwidth]{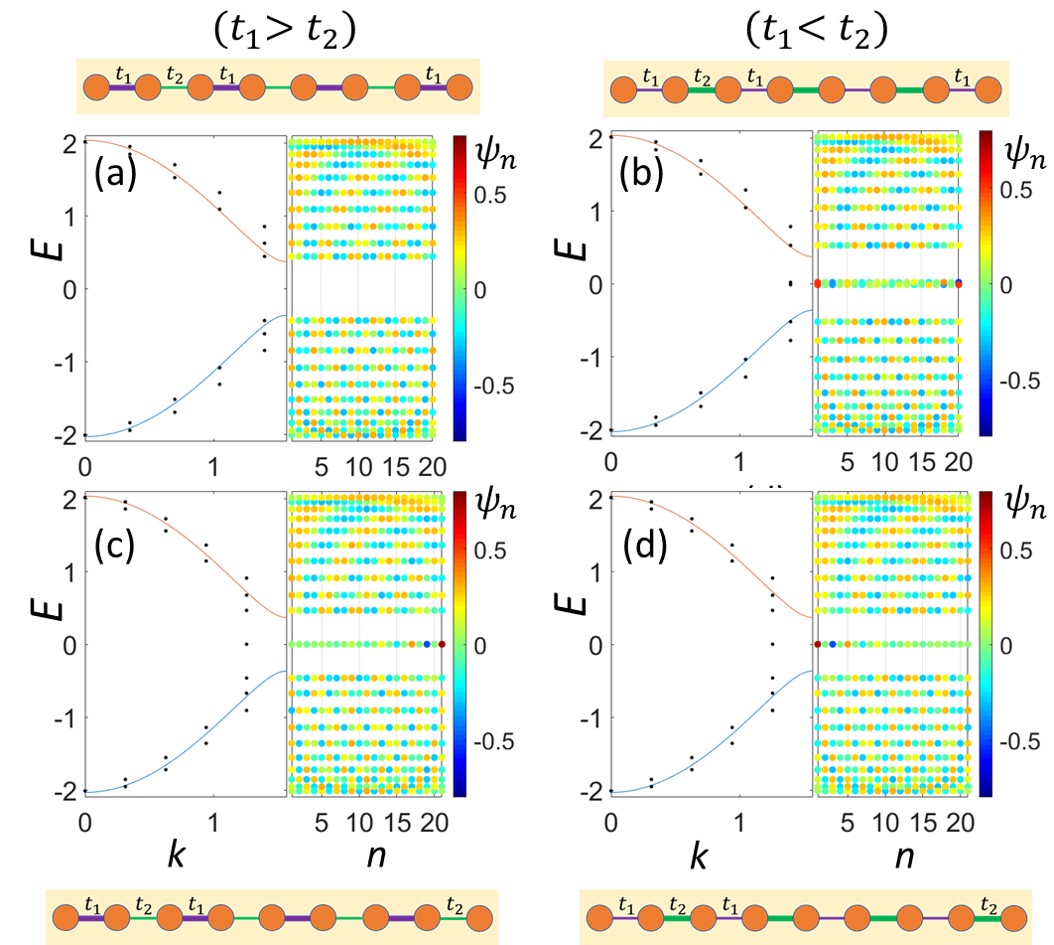}
	\caption{(color online) The 4 possible configurations of the finite SSH model: (a) and (b) correspond to even chain length ($N=20$), (a) with no edge states ($t_1>t_2$), (b) with 2 edge states  ($t_1<t_2$). (c) and (d) correspond to odd chain length ($N=21$), (c) with 1 edge state on the right ($t_1>t_2$) and (d) with 1 edge state on the left  ($t_1<t_2$). The dispersion curves for the infinite chain $N=\infty$ are shown as lines together with the discrete energy levels of the finite chain as dots. We used $t_1=1.2$ and $t_2=1/t_1$ on the left panels and the reverse for the right panels.
	}
	\label{sketcha}
\end{figure}

There are several noteworthy features of the SSH model that have attracted a substantial amount of interest in the literature. For instance, the infinite chain has a gap given by $2|t_1-t_2|$ \cite{PhysRevLett.42.1698} and illustrated in Fig.~\ref{sketcha}. The existence of this gap constitutes the cornerstone for the topological features of the SSH model. For periodic dimers coupled by $t_1$, the topological winding number (1D Berry phase or Zak phase \cite{zak1989berry}) is 1 if $t_1>t_2$ but zero for $t_1<t_2$. The reverse is true when considering dimers coupled by $t_2$. Hence, the infinite SSH model with $t_1\neq t_2$ exhibits two topologically distinct insulating phases depending on how the dimers are constructed \cite{ref-asboth}. In the presence of kink defects the model exhibits soliton solutions at zero energy with charge fractionalization \cite{jackiw1976solitons}. These zero-energy solutions are exponentially localized at the defects.

Similar exponentially localized solutions exist also in the finite SSH chain, but this time at the edge. The structure of these solutions is summarized in Fig.~\ref{sketcha} and will be discussed in more detail in the next section. In short, for energies near the band center there exists 0, 1 or 2 edge states depending on the parity of the site number and the values of the coupling strengths. Outside the band center, one recovers typical Bloch solutions as expected for periodic systems (see Fig.~\ref{sketcha}).

\subsection{Analytical solution of the SSH model}
Many features of the SSH hamiltonian can be determined analytically; we do so in some detail here for the case $N$ even (writing $N=2M$) mainly as a review and also to establish notation to be used in what follows. Many of the results are known (see, for instance, \cite{delplace2011zak,ref-asboth,PhysRevB.97.195439}). At the end of this section, the case $N$ odd will be discussed in much less detail.

The Schroedinger equation for a solution of energy $E$
\beq
\left( \hssh-E \right)\ket{\psi}=0
\label{eq-EqSch}
\eeq
couples any site with its immediate neighbours. Translational invariance (by an even number of sites) suggests the following ansatz for an eigenstate of $\hssh$:
\beq
\ket{\psi}=\sum_{n=0}^{M-1} \left( A\ket{2n+1}+B\ket{2n+2} \right) e^{in2k}.
\label{eq-psi1}
\eeq
The middle components (all but the first and last) of \eqref{eq-EqSch} reduce to a pair of equations:
\beq
\left( \begin{matrix}
	-E & t_1 + t_2 e^{-i2k} \\
	t_1 + t_2 e^{i2k} & -E
\end{matrix} \right)\left( \begin{matrix}
	A \\
	B
\end{matrix} \right) = 0.
\label{eq-middleeqns}
\eeq
A nontrivial solution requires that the determinant of the matrix vanish, giving the following dispersion relation:
\beq
E^2 = t_1^2 + t_2^2 + 2 \,t_1 t_2 \cos 2k.
\label{eq-Esq}
\eeq
If $(t_1-t_2)^2<E^2<(t_1+t_2)^2$, the wave number $k$ is real (corresponding to bulk states; this range gives the energy bands in the continuum limit), whereas outside this range it is complex (any such solution being an edge state, as we will see). For the moment, we will assume $k$ is real.

Since $k\to k+\pi$ has no effect on the ansatz, $k$ can be taken to be in the range $(-\pi/2,\pi/2]$. For any energy other than $E^2=(t_1\pm t_2)^2$ (a case which can safely be ignored), there are two solutions to \eqref{eq-Esq}, which we will write $\pm k$, where $0<k<\pi/2$.

It is useful to write $t_1+t_2 e^{\pm i2k}=|E|e^{\pm i2\varphi}$; like $k$, $0<\varphi<\pi/2$. Then the solution to \eqref{eq-middleeqns} can be written
\beq
\left( \begin{matrix}
	A \\
	B
\end{matrix} \right) =
\left( \begin{matrix}
	e^{-i\varphi} \\
	\pm  e^{i\varphi}
\end{matrix} \right),
\label{eq-AB}
\eeq
where here and in what follows the upper (lower) sign is for the solution of positive (negative) energy. The most general solution to the middle components of \eqref{eq-EqSch} is given by a sum of \eqref{eq-psi1} with \eqref{eq-AB} and the same expression with $(k,\varphi)\to(-k,-\varphi)$:
\begin{align}
\ket{\psi_\pm} = &\sum_{n=0}^{M-1} \Big\{
\big( C_+e^{-i\varphi} e^{i2nk} + C_-e^{i\varphi} e^{-i2nk} \big) \ket{2n+1} \nonumber\\
& \pm \big( C_+e^{i\varphi} e^{i2nk} + C_-e^{-i\varphi} e^{-i2nk} \big) \ket{2n+2} \Big\}
\label{eq-psi2}
\end{align}
where $C_\pm$ are constants.

The first and last components of \eqref{eq-EqSch} give
\beq
\left( \begin{matrix} e^{i\varphi} e^{-i2k} & e^{-i\varphi} e^{i2k}\\
	e^{-i\varphi} e^{iNk} & e^{i\varphi} e^{-iNk} \end{matrix} \right)
\left( \begin{matrix} C_+\\C_- \end{matrix} \right) = 0.
\label{eq-EdgeEqs}
\eeq
As above, a nontrivial solution requires that the determinant vanish, giving the following equation for $k$, written in terms of the hopping-parameter ratio $r \equiv t_1/t_2$ and $\text{s}_j \equiv \sin(j k)$:
\beq
r \, \text{s}_{N+2} + \text{s}_N = 0.
\label{eq-k}
\eeq
We will return to the solution of this equation, and the energy spectrum which follows from \eqref{eq-Esq}, shortly.

Solving \eqref{eq-EdgeEqs} for $C_\pm$ and substituting in \eqref{eq-psi2} gives
\begin{align}
\ket{\psi^\text{bulk}_\pm} = \sum_{n=0}^{M-1} \Big\{&
\sin\big[ (2n+2)k-2\varphi \big]\ket{2n+1}\nonumber\\
& \pm \sin\big[ (2n+2)k \big]\ket{2n+2} \Big\},
\label{eq-psi3}
\end{align}
where we have noted explicitly that these states, being oscillatory, are bulk states.
The coefficient of $\ket{2n+1}$ can be simplified slightly as follows. First, note that \eqref{eq-k} implies $\sin\big[ (N+2)k-2\varphi \big]=0$ which, given the range of $\varphi$, then implies $2\varphi = (N+2)k \mod \pi$. Now, modding by $\pi$ either has no effect on the coefficient or changes it by a sign, depending on whether the subtraction is an even or odd multiple of $\pi$, so the coefficient is, up to a sign, $\sin[(N-2n)k]$. This sign turns out to be that of $\sin(Nk)$ giving, finally,
\begin{align}
\ket{\psi^\text{bulk}_\pm} = \sum_{n=0}^{M-1} \Big\{&
\text{sgn}(\text{s}_N)\,\text{s}_{N-2n}\ket{2n+1}\nonumber\\
& \pm \text{s}_{2n+2}\ket{2n+2} \Big\}.
\label{eq-psi4}
\end{align}

Let us return now to the solution of \eqref{eq-k} for $k$. Recall that, if real, $k$ is in the range $[0,\pi/2]$. It can be shown that the endpoints (which correspond to $E^2=(t_1\pm t_2)^2$) can be excluded. Eq.~\eqref{eq-k} cannot be solved analytically for $k$, but it can be solved graphically. The number of solutions in the range $(0,\pi/2)$ depends, naturally, on the site number. More surprisingly, it also depends on the parameter $r$. This parameter has a critical value given by \cite{delplace2011zak}
\beq
\rc \equiv \frac{N}{N+2}.
\eeq

If $r>\rc$, there are $N/2$ solutions of \eqref{eq-k} for $k$ in $(0,\pi/2)$. Each pair $k,-k$ corresponds to two energy eigenstates of the form \eqref{eq-psi4} with equal and opposite energies. Thus, there are a total of $N$ solutions, forming a complete set of solutions of \eqref{eq-EqSch}. Since all these solutions have oscillatory behavior as a function of the site index, they are bulk states, as was mentioned earlier.

If, however, $r<\rc$, there is one fewer solution of \eqref{eq-k} in $(0,\pi/2)$, giving a total of $N-2$ solutions. But $\hssh$ clearly has $N$ eigenvalues and eigenvectors, so two have yet to be found. The missing solutions have complex wave numbers. If we substitute $k=\pi/2 + i \ka$ into \eqref{eq-k}, we find
\beq
\frac{\sinh(N\ka)}{\sinh((N+2)\ka)} = r.
\label{eq-kappa}
\eeq
This equation cannot be solved analytically, but it is easy to see that there are two real, equal and opposite solutions for $\ka$ if $r<\rc$ and none if $r>\rc$, which is exactly what is needed to make up for the two missing solutions of \eqref{eq-EqSch} for $k$ real. Defining $\ka$ to be the positive solution, the two corresponding solutions of \eqref{eq-EqSch}, identified with edge states since they have an exponential nature, turn out to be:
\begin{align}
\ket{\psi^\text{edge}_\pm} = \sum_{n=0}^{M-1} (-)^n\bigg\{&
\text{sh}_{N-2n}\ket{2n+1}\nonumber\\
& \pm \text{sh}_{2n+2}\ket{2n+2} \bigg\}
\label{eq-psi-edge},
\end{align}
where we have written $\text{sh}_j = \sinh(j \ka)$.

\begin{figure}[h!]
	\centering
	\hspace*{0cm} 
	\includegraphics[width=0.4\textwidth]{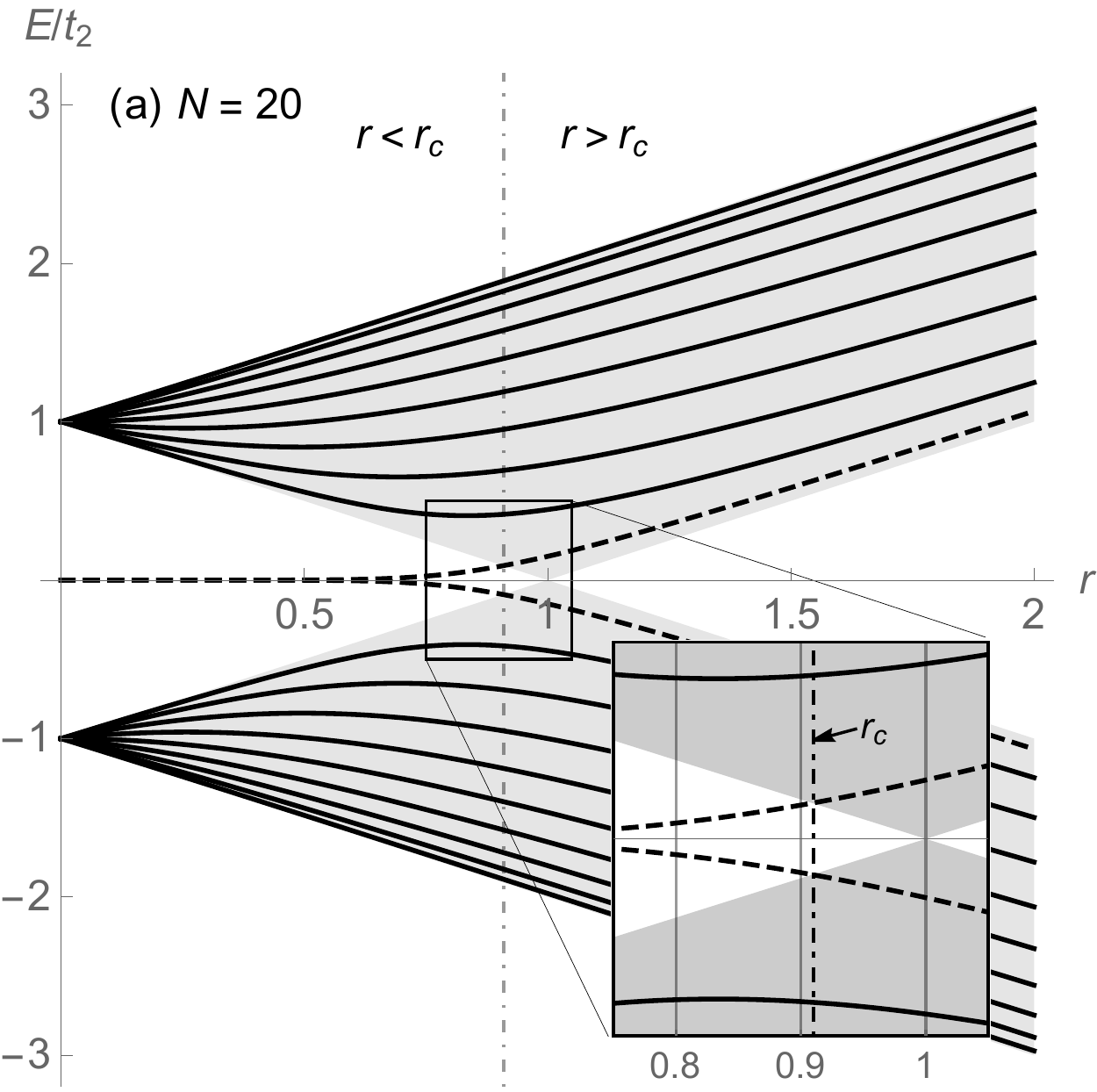}
	\\ \vspace{.2cm}
	\includegraphics[width=0.4\textwidth]{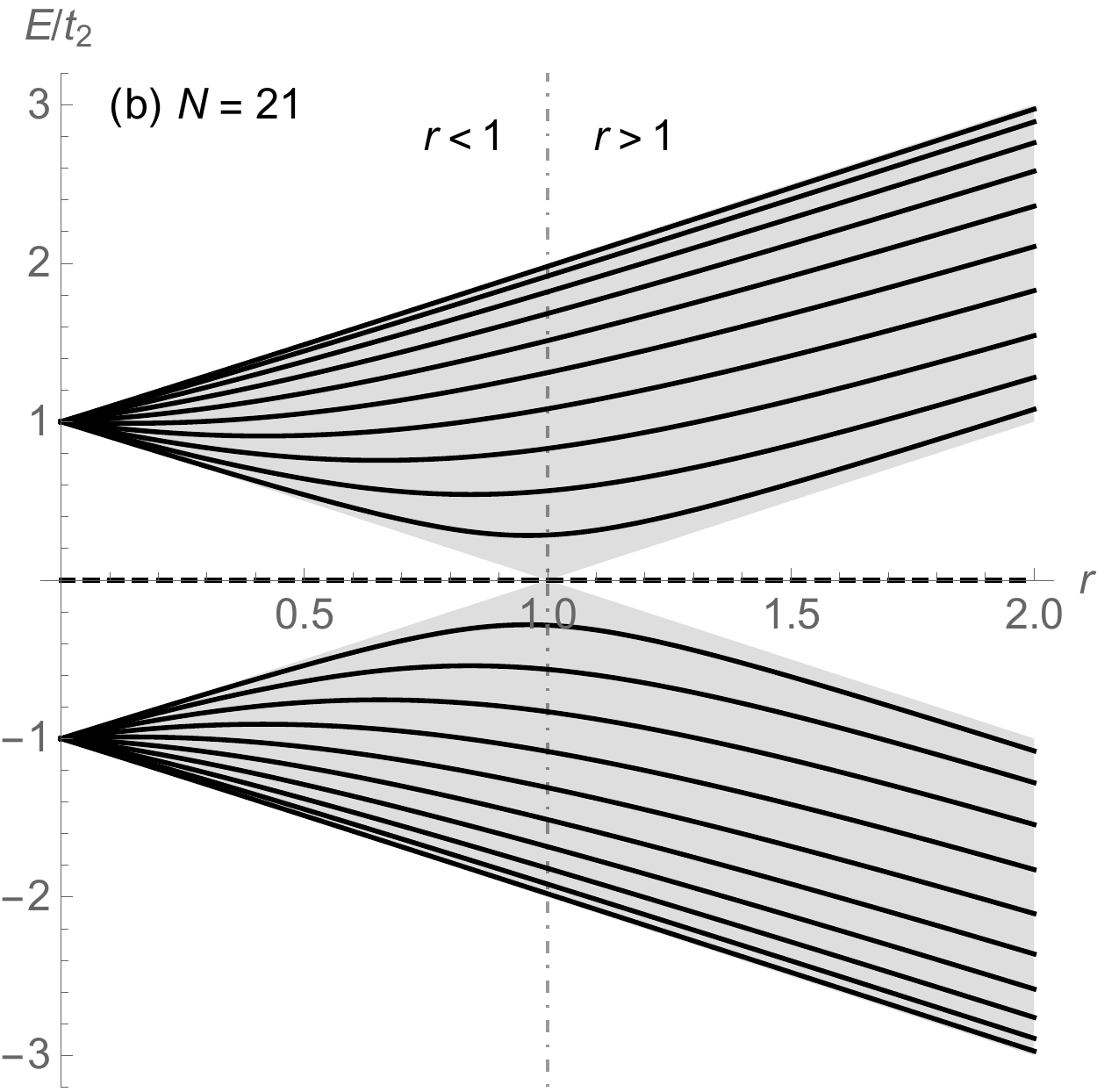}
	\caption{Energy spectrum of the SSH hamiltonian as a function of the hopping-parameter ratio $r=t_1/t_2$ for even and odd site numbers. The grey areas are the energy bands in the thermodynamic limit ($N\to\infty$). (a) $N=20$. The solid lines are bulk states for all $r$; the dashed lines are states which are bulk (edge) states for $r>\rc$ ($r<\rc$). Notice that (see inset) the edge states are exactly at the band edges for $r=\rc$. (b) $N=21$. The solid lines are bulk states for all $r$; the dashed line is the zero-mode which is a right (left) edge state for $r>1$ ($r<1$).}
	\label{Energy-N20N21}
\end{figure}

The wave numbers are determined by \eqref{eq-k} and, if $r<\rc$, \eqref{eq-kappa}. The energies are then determined by \eqref{eq-Esq}. The energy spectrum for $N=20$ is displayed as a function of $r$ in Fig.~\ref{Energy-N20N21} (a). (The energy spectrum has appeared in various forms in the literature; see for instance \cite{PhysRevLett.100.096407,delplace2011zak,ref-asboth,ref-li,ref-gu,PhysRevB.96.205424,ref-batra}.) The most striking feature is the existence of states in the band gap (defined in the thermodynamic limit) for $r<\rc$. These are the edge states given by \eqref{eq-psi-edge}.

\begin{figure}[h!]
	\centering
	\hspace*{-0.2cm}\includegraphics[width=0.5\textwidth]{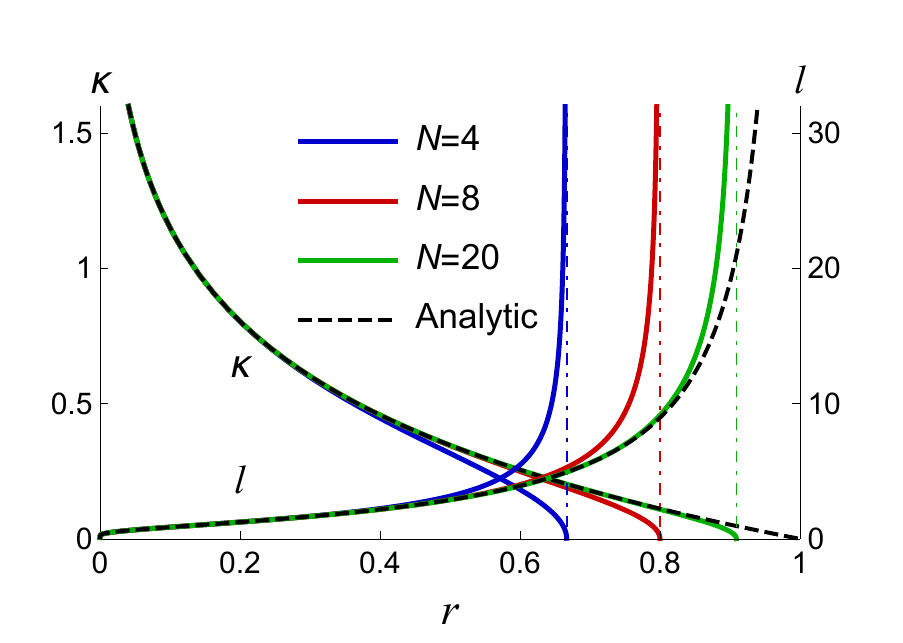}
	\caption{(color online) Decay parameter $\ka$ and penetration length $l=1/\ka$ for edge states as determined by \eqref{eq-kappa}, displayed as a function of the hopping-parameter ratio $r=t_1/t_2$ for several values of $N$. There is no solution (and therefore there are no edge states) if $r>\rc=N/(N+2)$ (vertical broken lines).
	 Also shown (dashed lines) is the analytic approximation given in \eqref{eq-kappa-2}; only the $N$-independent first term is included. This is also the penetration length for any $N$ odd (see \eqref{eq-penlengthnodd}). }
	\label{fig-kappa}
\end{figure}

Describing the states given by \eqref{eq-psi-edge} as {\em edge states} merits some discussion. On the one hand, the wave number is complex, so for a sufficiently large system the states are confined to the edges with a penetration length into the bulk given by $l\equiv1/\ka$. On the other hand, as $r \to \rc$ from below, $\ka$ goes to zero and the penetration length goes to infinity (see Fig.~\ref{fig-kappa}). Thus for any finite-size system and $r$ sufficiently close to $\rc$, the penetration length is longer than the system size and the state is for all intents and purposes no longer confined to the edges, rendering it relatively indistinguishable from the rest of the states.

An approximate analytic solution to \eqref{eq-kappa} can be given if $N$ is large and/or $r$ is small. One finds
\beq
\ka = \frac{|\log r|}{2} - \frac{r^N(1-r^2)}{2} + O(r^{2N}).
\label{eq-kappa-2}
\eeq
The first term becomes dominant rapidly as either $N$ gets large or $r$ gets small; in Fig.~\ref{fig-kappa} only that term is included in the analytic curves.

We can derive an approximate analytic expression for the edge-state energies by substituting $k=\pi/2 + i \ka$ into \eqref{eq-Esq}, with $\ka$ given by \eqref{eq-kappa-2}. The dominant term in \eqref{eq-kappa-2} gives $E=0$ (and indeed these states are often erroneously described as zero-energy states); the energies are dominated by the second term, giving
\beq
E=\pm t_2 \,r^{N/2} (1-r^2) + O(r^{N}),
\label{eq-edgestateenergies}
\eeq
showing the well-known exponentially decreasing behavior of the energies as a function of $N$ \cite{delplace2011zak,ref-asboth}.

\begin{figure}[h!]
	\centering
	\hspace*{-0.3cm}\includegraphics[width=0.5\textwidth]{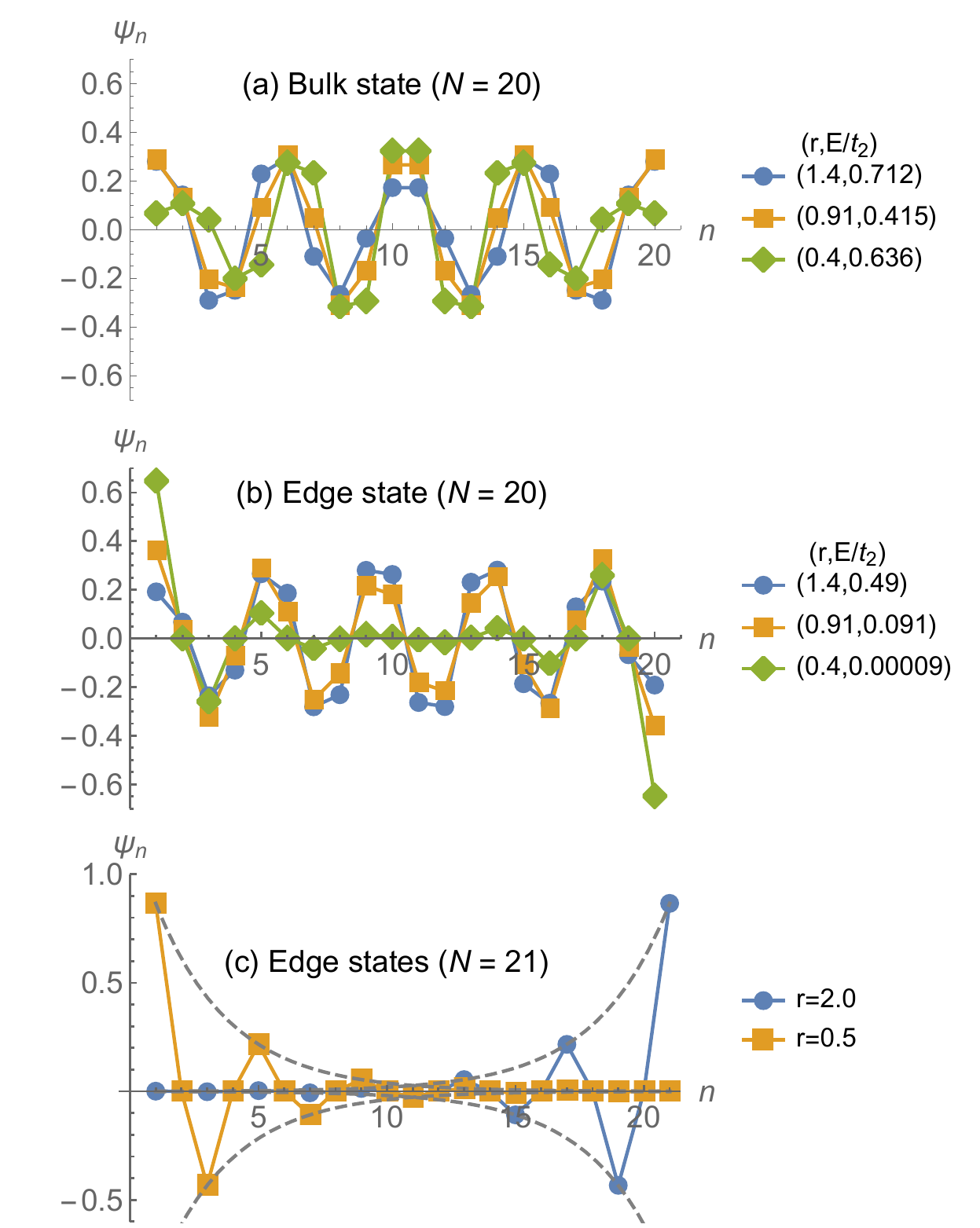}
	\caption{(color online) Examples of SSH bulk and edge states. (a) Bulk state (given by \eqref{eq-psi4}), for $N=20$ and three values of the hopping-parameter ratio $r$, the middle of which is $\rc$. The state chosen is the lowest positive-energy solid line in Fig.~\ref{Energy-N20N21} (a). Key observation: all states are large throughout the system, as expected given the oscillatory nature of the solution \eqref{eq-psi4}. (b) State that transitions to an edge state (given by \eqref{eq-psi-edge} below $\rc$) as $r$ is decreased, for $N=20$ and three values of the hopping-parameter ratio $r$, the middle of which is $\rc$. The state chosen is the positive-energy dashed line in Fig.~\ref{Energy-N20N21} (a). Key observation: the first two states are described by \eqref{eq-psi4} and are large throughout the system, whereas the third is described by \eqref{eq-psi-edge} and decreases exponentially in the bulk. (c) Zero-energy edge state (right for $r>1$, left for $r<1$) for $N=21$ corresponding to the dashed line in Fig.~\ref{Energy-N20N21} (b). The dotted lines are the exponential envelope functions $\pm \exp[(n/2)\log r]$ (up to normalization).}
	\label{fig-allstates}
\end{figure}

The distinction between bulk and edge states, and just how ``edgy" the edge states are, is illustrated in Fig.~\ref{fig-allstates} (a,b), showing a bulk state for all $r$ and a state whose nature (edge \vs bulk) changes at $r=\rc$, respectively.

We conclude this section with a brief discussion of the case $N$ odd (writing $N=2M+1$). Much of the above analysis applies with only slight modification. It turns out that for all values of $r$ there are $N-1$ bulk states with equal and opposite energies and one zero-energy state. The latter is localized on the left (right) edge for $r<1$ ($r>1$). This state is easily constructed by noting that \eqref{eq-EqSch} with $E=0$ decouples into two sets of equations: one for the coefficients of odd sites and one for those of even sites. The latter are zero, and it is easy to show that the (unnormalized) zero-energy state is
\beq
\ket{\psi_0}=\sum_{n=0}^M (-r)^{n}\ket{2n+1}.
\label{eq-zeromode}
\eeq
Clearly if $r>1$ the state grows exponentially from left to right with the opposite conclusion if $r<1$, with penetration length
\beq
l = 2/|\log r| \qquad (N~\text{odd})
\label{eq-penlengthnodd}
\eeq

Unlike the case $N$ even, for which the penetration length is given by the inverse of \eqref{eq-kappa-2}, for $N$ odd the penetration length is independent of $N$. The two agree in the large-$N$ limit, as expected.
The energy spectrum as a function of $r$ is displayed in Fig.~\ref{Energy-N20N21} (b) and the edge state for two representative values of $r$ is shown in Fig.~\ref{fig-allstates} (c).

Having established the main properties of the SSH chain, we now consider the coupling of a qubit to the SSH chain.


\section{SSH chain coupled to semi-infinite lead and qubit}

\begin{figure}[h!]
	\includegraphics[width=0.48\textwidth]{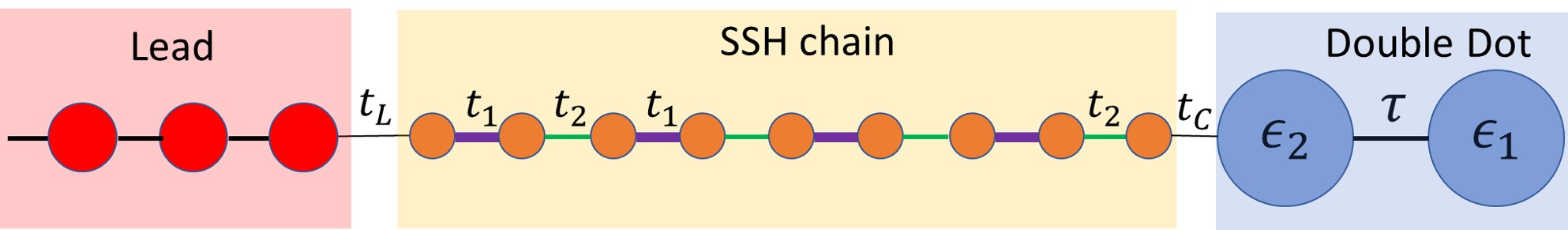}
	\caption{Full system: SSH chain coupled to a semi-infinite lead on one side and a qubit on the other side. (The example shown has $N$ odd, so the rightmost SSH hopping parameter is $t_2$; if $N$ is even, it would be $t_1$.)}
	\label{fig-FullSys}
\end{figure}

In this section we will consider the full system consisting of the SSH chain coupled on the left to a semi-infinite lead and on the right to a qubit (Fig.~\ref{fig-FullSys}). Since we will ultimately incorporate the effects of the SSH chain and semi-infinite lead into an effective qubit Hamiltonian, in the next subsection we will quickly review the relevant features of the qubit. We will then study the full system in the following subsection.

\subsection{Isolated qubit}

The isolated qubit, or double dot (rightmost subsystem in Fig.~\ref{fig-FullSys}), is described by the Hamiltonian
\beq
\hdd=\left( \begin{matrix}
	\eps_2 & \tau\\
	\tau & \eps_1
\end{matrix} \right).
\label{eq-H_DD}
\eeq
The (energy-dependent) Green's function, defined by $G^\text{DD} = (E-\hdd)^{-1}$, is easily calculated; for instance, its (1,2) component is
\beq
G^\text{DD}_{12}(E) = \frac{\tau}{\delta}\left(\frac{1}{E-\la_+ + i0^+}-\frac{1}{E-\la_- + i0^+}\right),
\label{eq-GDD12}
\eeq
where
\bea
\delta&=&\sqrt{(\eps_1-\eps_2)^2+4\tau^2}=\sqrt{{\de_0}^2+4\tau^2},\nonumber\\
\la_\pm&=&\frac{1}{2}(\eps_1+\eps_2 \pm\de) = \epsbar \pm \frac{\de}{2}
\label{eq-del-lapm}
\eea
are the energy splitting and the energies of the full Hamiltonian, respectively,
and we have defined the uncoupled ($\tau=0$) energy splitting $\de_0=\eps_1-\eps_2$ and the average energy $\epsbar = (\eps_1+\eps_2)/2$.

The infinitesimal positive quantity $0^+$ in \eqref{eq-GDD12} gives the pole prescription necessary to compute the retarded time-dependent Green's function. This is obtained by Fourier transformation, giving zero for $t<0$ while for $t>0$ $G^\text{DD}(t)$ consists of two terms oscillating at frequencies $\la_\pm$; for instance,
\begin{align}
\label{G12t-isolated}
G^\text{DD}_{12}(t)&=\int_{-\infty}^\infty dE\, e^{-iEt}G^\text{DD}_{12}(E)\nonumber\\
&=-\frac{2\pi i\tau}{\de}\left(e^{-i\la_+t}-e^{-i\la_-t}\right)
\nonumber\\
&=-\frac{4\pi \tau}{\de}e^{-i\epsbar t}\sin(\de t/2).
\end{align}

For an isolated qubit the time dependent solutions are simply periodic. However, if the qubit is coupled to an external infinite system (for example, a semi-infinte lead), the oscillations will decay. The decoherence of a qubit can then be evaluated by evaluating the off-diagonal element of the Green's function \cite{eleuch2017probing}. The coupling between the qubit and external system can take on different forms. Here we restrict ourselves to the simplest case, where the qubit is a double dot connected to an external system via a small tunneling coupling.

\subsection{Double dot coupled to SSH chain and semi-infinite lead}

As was mentioned in the introduction, our main interest is to investigate how the dynamics of a double dot is affected by the presence of edge states in an adjoining system. We therefore consider a system composed of three parts: the double dot coupled to one end of an SSH chain which is connected at the other end to a semi-infinite lead. The Hamiltonian is
\beq
H=\left(\begin{array}{c|c|c}
\hinf & W & 0\\
 \hline
W^\dagger & \hssh & V_N\\
\hline
0 & V_N^\dagger & \hdd
\end{array}\right)
\label{H-dot+chain+lead}
\eeq
where $\hdd$ and $\hssh$ are defined in \eqref{eq-H_DD} and \eqref{eq-H_SSH}, $V_N$ is a $N\times 2$ matrix whose only nonzero element is $(V_N)_{N1}=\tc$, $W$ is an $\infty \times N$ matrix whose only nonzero element is $W_{\infty 1}=\tl$, and $\hinf$ is
\beq
\hinf=\left(\begin{matrix}
0 & 1 \\
1 & 0 & 1 & \phantom{\ddots} \\
& 1 & 0 & \ddots \\
\phantom{\ddots} & \phantom{\ddots} & \ddots & \ddots
\end{matrix}\right),
\label{eq-H_infty}
\eeq

The dynamics of the entire system can be determined by evaluating its Green's function. But since we are interested only in that of the double dot, we can incorporate the effect of the SSH chain and lead in a self-energy using standard techniques (see for instance \cite{datta2005quantum} for a general discussion; we will adapt the analysis of \cite{eleuch2017probing}, following the notation introduced there, to the current system).

It is useful to do this in two steps. First, the effect of the lead on the SSH chain can be incorporated by replacing the lead by a modification of $\hssh$:
\begin{align}
\left(\begin{array}{c|c}
\hinf & W\\
\hline
W^\dagger & \hssh
\end{array}\right)
&\to
\left( \begin{matrix}
	\Sii      & t_1    & 0      & \cdots & 0      \\
	t_1    & 0      & t_2    & \ddots & \vdots \\
	0      & t_2    & \ddots & \ddots & 0      \\
	\vdots & \ddots & \ddots & 0      & t      \\
	0      & \cdots & 0      & t      & 0 \\
\end{matrix} \right) \nonumber\\
&\equiv \hsshi.
\label{eq-hsshi}
\end{align}
Thus, $\hsshi$ is identical to $\hssh$ except for the upper-left element, which is a self-energy proportional to the surface Green's function of the lead \cite{eleuch2017probing}:
\beq
\Sii = \tl^2 G_\infty^\text{S}(E)
=\frac{\tl^2}{2}\left(E-i\sqrt{4-E^2}\right).
\label{eq-Siginfp}
\eeq

The second step repeats the above, incorporating the effect of $\hsshi$ on the double dot by an appropriate modification of $\hdd$:
\begin{align}
\left(\begin{array}{c|c}
\hsshi & V_N\\
\hline
V_N^\dagger & \hdd
\end{array}\right)
&\to
\left( \begin{matrix}
\eps_2 + \Sisshi & \tau\\
\tau & \eps_1
\end{matrix} \right) \nonumber\\
&\equiv \hdd_{\text{SSH},\infty}.
\label{eq-hddsshi}
\end{align}
As above, the only modification of $\hdd$ needed is a term added to the upper-left element. This added term is proportional to the surface Green's function of $\hsshi$:
\beq
\Sisshi = \tc^2 \GSsshi(E)
\label{eq-sisshi}
\eeq
The surface Green's function $\GSsshi$
is the $(N,N)$ component of the Green's function of $\hsshi$ defined by
\beq
\Gsshi(E-\hsshi) = \mathbb{1}.
\label{eq-Gsshi}
\eeq
To determine $\GSsshi$, we write the first row of $\Gsshi$ in a form identical to \eqref{eq-psi1} and follow the steps used to derive \eqref{eq-psi3}. (The current calculation is somewhat easier because we only need the last component.) Here as above, the cases $N$ even and odd must be handled differently. We find
\ignorethis{
\beq
\GSsshi =
\begin{cases}
	\frac{Et_2{s}_N-\Sigma_{\infty}(t_1{s}_{N-2}+t_2{s}_N)}{t_2^2(t_1{s}_{N+2}+t_2{s}_N)-Et_2\Sigma_{\infty}{s}_N}, & (N\text{ even})\\
	\frac{t_1(t_1{s}_{N-1}+t_2{s}_{N+1})-E\Sigma_{\infty}{s}_{N-1}}{t_1t_2E{s}_{N+1}-t_2\Sigma_{\infty}(t_1{s}_{N+1}+t_2{s}_{N-1})}, & (N\text{ odd})
\end{cases}
\label{Green_Analytical}
\eeq
}
\beq
\GSsshi =
	\frac{Et_2{s}_N-\Sigma_{\infty}(t_1{s}_{N-2}+t_2{s}_N)}{t_2^2(t_1{s}_{N+2}+t_2{s}_N)-Et_2\Sigma_{\infty}{s}_N}
\label{Green_Analytical-even}
\eeq
if $N$ is even and
\beq
\GSsshi =
	\frac{t_2(t_2{s}_{N-1}+t_1{s}_{N+1})-E\Sigma_{\infty}{s}_{N-1}}{t_1t_2E{s}_{N+1}-t_1\Sigma_{\infty}(t_2{s}_{N+1}+t_1{s}_{N-1})}
\label{Green_Analytical-odd}
\eeq
if $N$ is odd.

\begin{figure}[h!]
	\centering
	\hspace*{0cm}\includegraphics[width=0.47\textwidth]{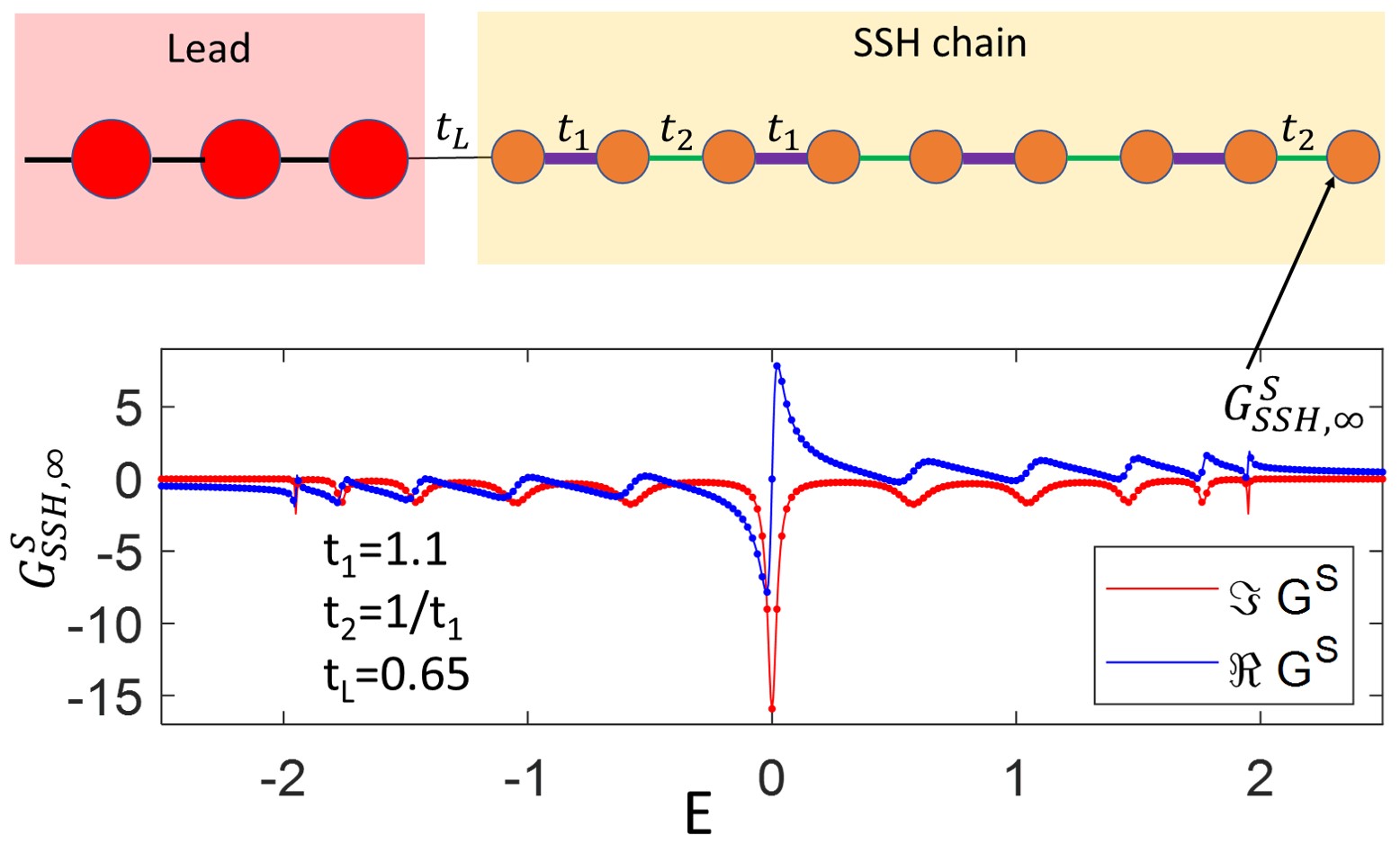}
	\caption{(color online) Line plots of the surface Green's function  $\GSsshi$ \eqref{Green_Analytical-odd} of the right most site as a function of energy, while the dots are obtained by evaluating numerically the full matrix \eqref{eq-hsshi}. Here $N=11$ and $t_1>t_2$ were used (corresponding to case (c) in Fig.~\ref{sketcha}).}
	\label{sketchb}
\end{figure}

The surface Green's functions is shown in Fig.~\ref{sketchb} for the case where $N$ is odd, where there is an edge state on the right of the SSH chain. This corresponds to case (d) shown in Fig.~\ref{sketcha}. The striking feature is the large negative imaginary part of the Green's function at zero energy, corresponding to the large local density of states of the right edge state. This large negative imaginary part of the surface Green's function is the main source of decoherence for the double dot coupled to the rightmost site of the SSH chain.

The energy-dependent Green's function for the double dot including the effect of the SSH chain and lead is obtained in a straightforward manner by the substitution
\beq
\eps_2 \to \eps_2 + \Sisshi \equiv \eps_2'
\label{eq-eps2sub}
\eeq
in \eqref{eq-GDD12} (and similar equations for the other components). It is useful to define
\begin{align}
\de' &= \left.\de\right|_{\eps_2 \to \eps_2'} = \sqrt{(\eps_1-\eps_2')^2+4\tau^2}
\label{eq-dep}\\
\la_\pm' &= \left.\la_\pm\right|_{\eps_2 \to \eps_2'}
 = \epsbar + \frac{1}{2}(\Sisshi\pm \de').
 \label{eq-lap}
\end{align}

Since $\hsshi$ is not Hermitian (reflecting the fact that the double dot is not a closed system), the time-dependent Green's function will have decaying behavior (in contrast with the oscillatory behavior exhibited in \eqref{G12t-isolated}) from which the decoherence rate can be extracted by determining the slowest decay. This determines the long-term behavior of the double dot.

However, the time-dependent Green's function cannot be evaluated exactly since the very complicated dependence of $\Sisshi$ on $E$ precludes an exact evaluation of the Fourier transform of $\GSsshi$.

An analytic approximation can be obtained by noting that the frequencies in the isolated double dot Green's function $G^\text{DD}$ are $\la_\pm$. Eq.~\eqref{eq-lap} suggests using $\la_\pm'$ instead. This is not quite correct, since $\la_\pm'$ are energy-dependent. However, for small coupling between the double dot and the rest of the system, one can show \cite{eleuch2017probing} that to a good approximation the (now complex) frequencies should be evaluated at the corresponding poles of the isolated double dot Green's function, $\la_+'(\la_+)$ and $\la_-'(\la_-)$. According to this analytic approximation, the decay rates are given by the imaginary part of the frequencies, and we conclude that the decoherence time $\tau_\phi$ is given by
\beq
\left(\tau_\phi\right)^{-1} \approx \min \left( -\frac{1}{2}\Im\left\{\Sisshi(\la_\pm)
\pm\de'(\la_\pm)\right\} \right).
\label{coupletolead2}
\eeq
This expression makes it clear that there are two contributors to the decay rate, corresponding to the coupling of each double dot state to the SSH chain. The slower of the two rates dominates at long times, so it is this that determines $\left(\tau_\phi\right)^{-1}$.

Alternatively, one can determine the decoherence time by numerically evaluating the Fourier transform of the Green's function and extracting the decay constant of the long-time behavior. All matrix elements will decay with the same rate. Here we use the off-diagonal element to compute it. For the Fourier transform, it is important to use a very small discretization of the energy and we used $\Delta E\sim 10^{-5}$ (for a bandwidth of 4). The time dependence is then evaluated using a fast Fourier transform, which is fitted to multiple exponential decay functions, from which the slowest decay is extracted at long times.

\begin{figure}[h!]
	\centering
	\hspace*{0cm}\includegraphics[width=0.47\textwidth]{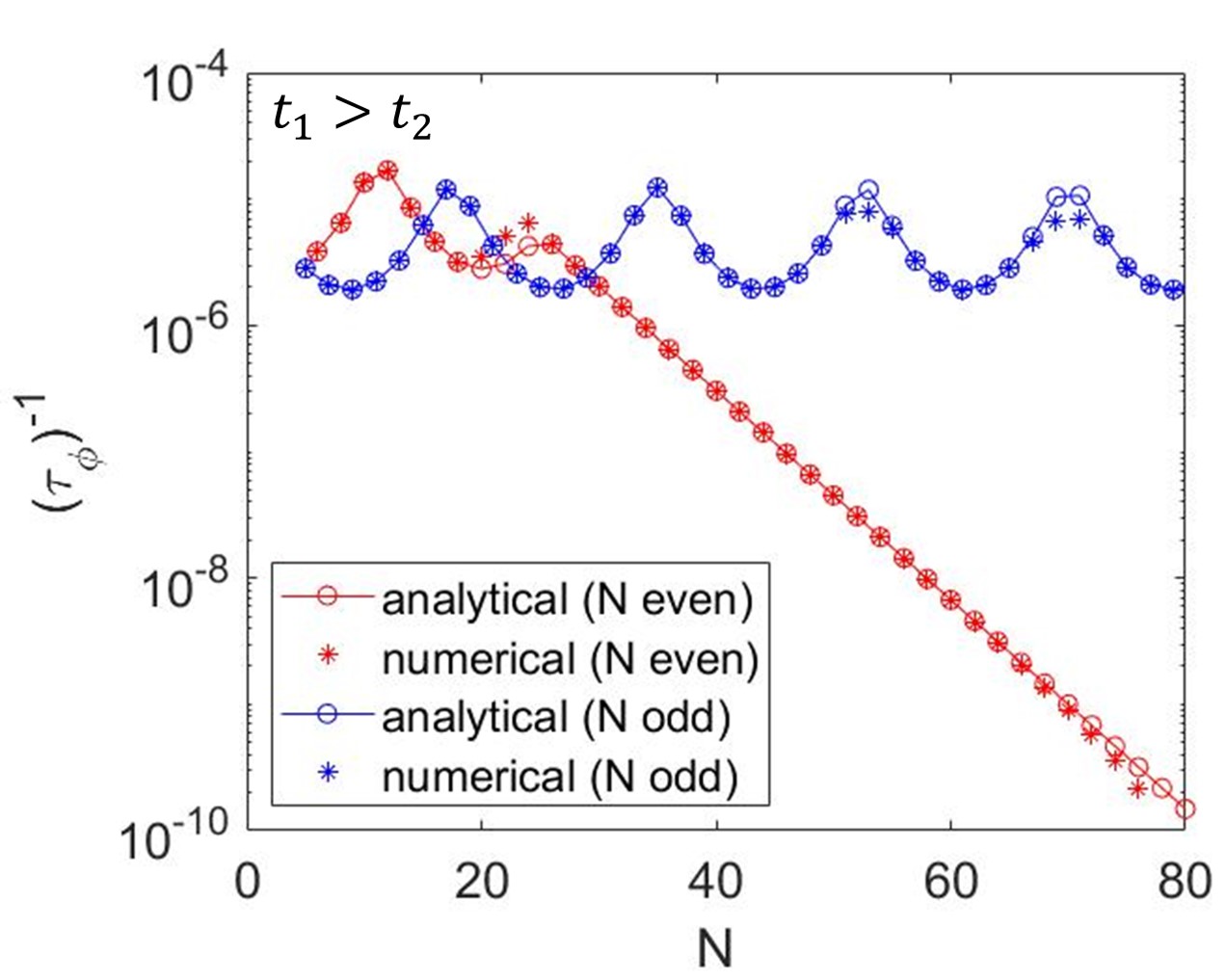}
	\caption{(color online) Decoherence rate for site number $N$ even (red) vs.~odd (blue) for $(t_1,t_2)=(1.1,1/1.1)$. The analytical expression uses (\ref{Green_Analytical-even},\ref{Green_Analytical-odd}) and \eqref{coupletolead2}, while the numerical computation fits the time-dependent Green's function evaluated via Fast Fourier Transform to an exponential function. The other parameter values are: $t_C=.035$, $\tau=.03$, $t_L=.65$, and double dot energies  $(\eps_1,\eps_2)=(0.4022,0.0022)$, chosen to give one zero eigenvalue for the isolated double dot: $\la_-=0$. The nonzero double dot energy is $\la_+=0.4044$, lying in the SSH energy band.}
	\label{fig-decoherence}
\end{figure}

\begin{figure}[h!]
	\centering
	\hspace*{0cm}\includegraphics[width=0.47\textwidth]{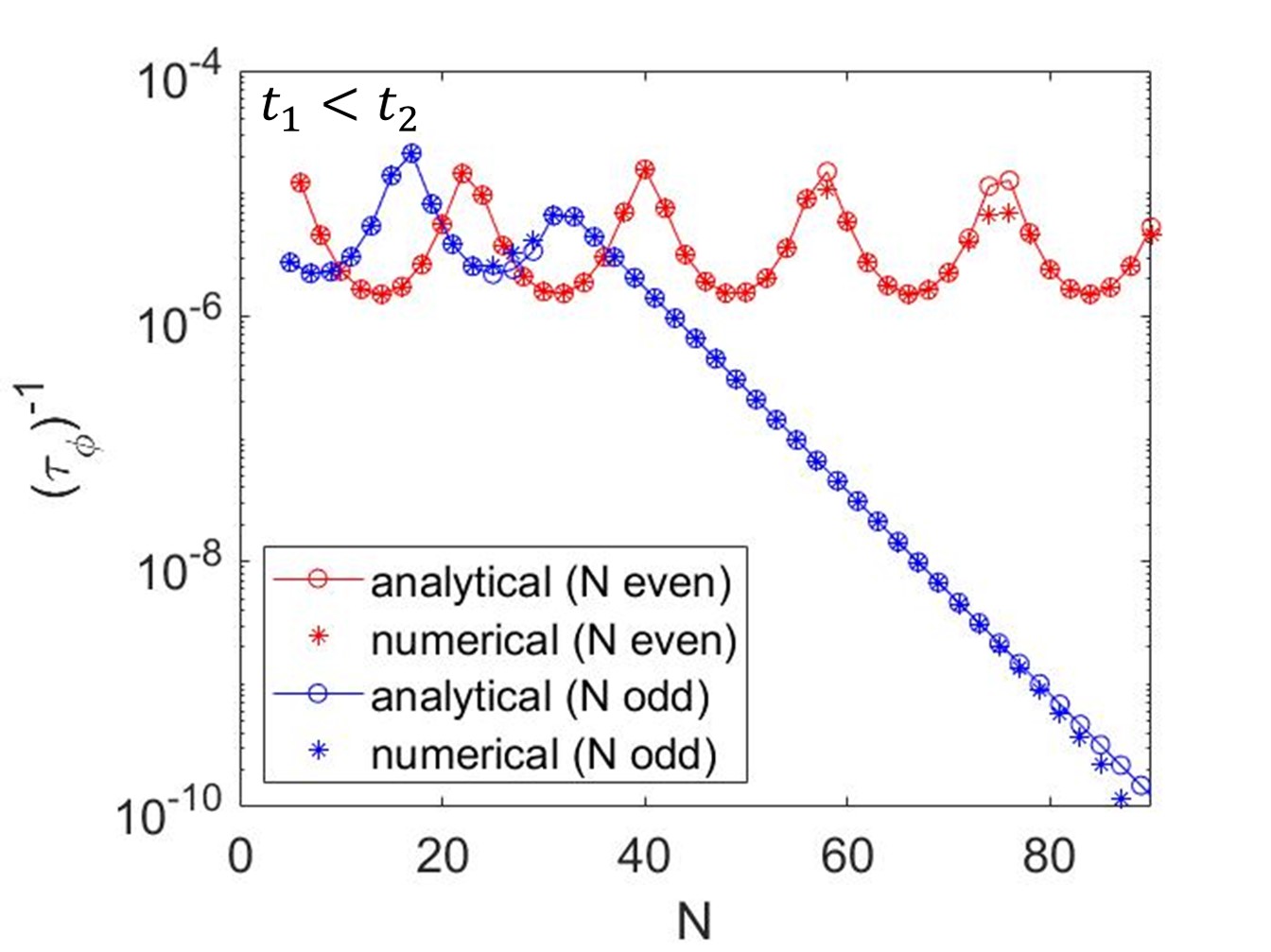}
	\caption{(color online) Decoherence rate for site number $N$ even (red) vs.~odd (blue) for $(t_1,t_2)=(1/1.1,1.1)$. Computational details and other parameter values are as in Fig.~\ref{fig-decoherence}.}
	\label{fig-decoherence2}
\end{figure}

Both methods will be used in what follows; the excellent agreement between the analytic approximation and numerical evaluation of the decoherence time is a convincing post hoc justification of the analytic approximation (see Figs.~\ref{fig-decoherence} and \ref{fig-decoherence2}). The parameters in these figures were chosen so that one double dot energy is zero (so that it couples to any SSH edge states, whose energies are also zero or exponentially small), while the other double dot energy lies in the SSH continuum. The latter gives rise to rapid decoherence, so the overall decoherence is determined by whether or not the zero-energy state also decoheres rapidly. In the discussion that follows, we therefore focus on the decoherence of the zero-energy state; we will see that this depends strongly on whether or not there is a same-side edge state.

There are some small deviations between the numerics and analytical solution \eqref{coupletolead2}. Most of the small differences can be attributed to extracting numerically the decay rates from a finite time interval of a strongly oscillating function.

While we have shown that the coupling of the qubit to the SSH chain affects its dynamics, the reverse is also true. The topological nature of the edge states in the SSH chain is also perturbed. However, the coupling of the qubit induces a perturbation of the SSH states of order $t_c^2$, which can be neglected for $t_c\ll 1$, which is the situation we are considering here.

We have seen above that there is an edge state of energy zero at the right edge of the SSH chain whenever $r>1$ if $N$ is odd, while if $r>\rc$ (a weaker condition) there are no states near $E=0$ (a gap) if $N$ is even. Both conditions are satisfied if $r>1$, in which case for a qubit coupled to the right edge of the SSH chain (with coupling $t_C$), the qubit will decohere much faster when there is an edge state ($N$ odd), while the decoherence rate will be exponentially suppressed for the gapped $N$ even case. This is what we see in Fig.~\ref{fig-decoherence}. Already for $N=80$, with the choice of parameters used in Fig.~\ref{fig-decoherence}, we see approximately a five-decade difference in the decoherence rate between the odd (one edge state) and the even (no edge states) cases. Hence, the qubit can act as a very sensitive detector of the edge state.

Rather different behavior occurs when $r<\rc$. In this case, when $N$ is odd there is an edge state on the left edge of the SSH chain, while for $N$ even there are two edge states, each confined to both edges. If we look at the decoherence rate of the qubit attached to the rightmost site (see Fig.~\ref{fig-decoherence2}), we observe an exponentially decaying rate for $N$ odd, but not for $N$ even. The even case is quite intuitive, since the existence of edge states on both sides of the SSH chain will naturally lead to an increased decoherence of the qubit. The odd case is different since there is no edge state to the right. However, despite having an edge state next to the lead, the decoherence rate is exponentially suppressed with $N$. This is because the local density of states at zero energy is close to zero at the right edge where the qubit is located, and therefore it does not lead to an increase in decoherence. 

For all cases, the system displays oscillatory behavior of the decoherence rate as a function of $N$ at small enough $N$. This is a consequence of resonances within the SSH chain: for values of $N$ for which an SSH energy is close to $\la_+$, the corresponding double dot state couples more strongly to the lead via the SSH chain, causing a higher decoherence rate compared to other values of $N$. When reducing $t_L$, the coupling to the lead, these resonances become more pronounced. For larger $N$ this behavior is then taken over by exponential decay as a function of $N$ when there is no edge state ($r>\rc$ and $N$ even) or when the edge state is on the opposite side of the qubit ($r<\rc$ and $N$ odd). This change of behavior can be understood mathematically from expression \eqref{coupletolead2}. For $N$ small, the minimum is given by the expression evaluated at the double dot eigenvalue within the SSH band ($\lambda_+$), while for $N$ large enough and when there are no edge states ($N$ even) or an edge state opposite to the qubit ($N$ odd) the minimum is given by the expression evaluated at the double dot eigenvalue at zero energy ($\lambda_-=0$). Alternatively, if there are two edge states ($N$ even) or if there is a single edge state next to the qubit ($N$ odd), the decoherence rate is determined by \eqref{coupletolead2} evaluated at $\lambda_+$. Evaluating \eqref{coupletolead2} at $\lambda_-=0$ would lead to a divergence with large $N$. Hence the minimum (and the physical decoherence rate) is determined by $\lambda_+$. Physically, this can be understood by identifying the decoherence rate of the qubit with the smallest escape rate from the double dot into the SSH chain.

\section{Discussion and conclusions}

\begin{figure}[h!]
	\centering
	\hspace*{0cm}\includegraphics[width=0.48\textwidth]{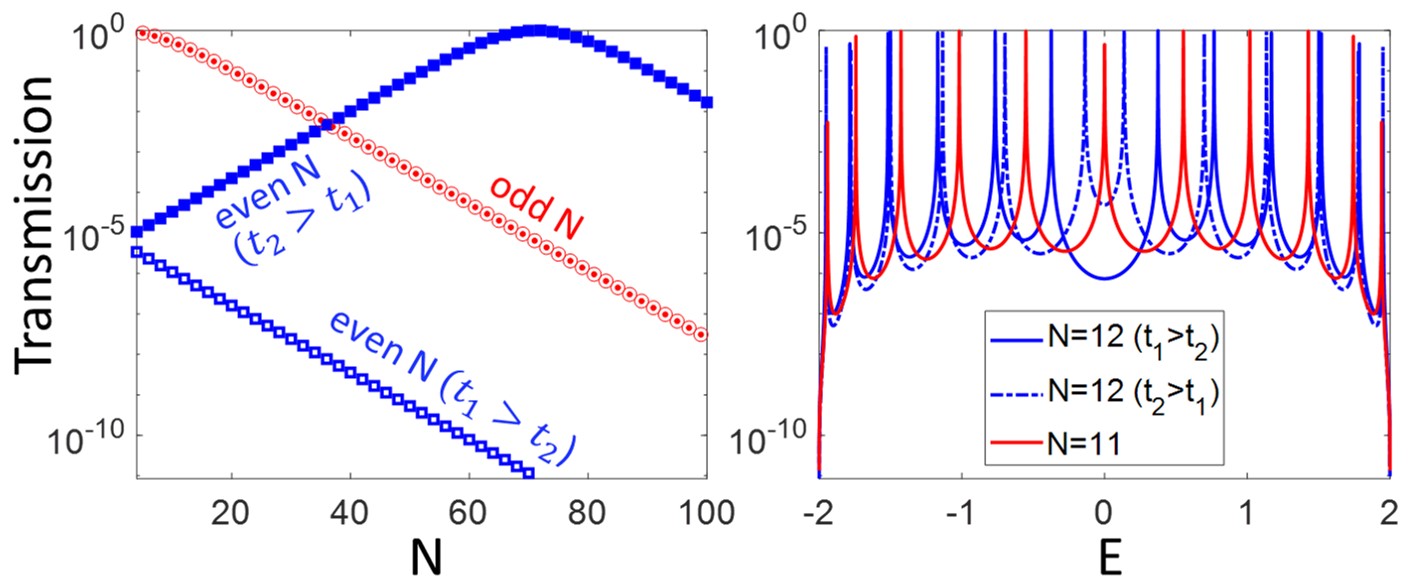}
	\caption{(color online) The transmission through a SSH chain coupled to 2 semi-infinite leads on both sides. Left: Transmission at $E=0$ as a function of $N$. Right: Transmission as a function of energy. The parameters used here are: $t_c=.035$ (the coupling between the SSH chain and the leads). $t_2=1.1$ and $t_1=1/t_2$ for the open squares on the left and broken line on the right.  $t_1=1.1$ and $t_2=1/t_1$ for the filled squares on the left and solid blue line on the right. For $N$ odd, the cases $t_2=1.1$, $t_1=1/t_2$ and $t_1=1.1$, $t_2=1/t_1$ have the same transmissions (shown in red open and closed circles).}
	\label{transmission}
\end{figure}

Interestingly, the exponential decrease in the rate of decoherence is similar for the case where we have an edge state on the opposite side of the qubit ($N$ odd in Fig.~\ref{fig-decoherence2}) and when there are no edge states at all ($N$ even in Fig.~\ref{fig-decoherence}). In both cases, the decoherence rate is exponentially suppressed. This is quite similar to the case of localization with a random potential along the chain, which also leads to an exponential suppression of the decoherence rate with chain length \cite{eleuch2017probing}. From this perspective, the case of a localized edge state close to the qubit is quite different, since despite it being a localized state, the decoherence rate is enhanced by the existence of a localized edge state, even though transmission \cite{ruocco2017transport,bohling2018thermoelectric} through the SSH chain will eventually be suppressed (see Fig.~\ref{transmission}). 

It is instructive to look at the case of transmission through the SSH chain as a reference. To this end, we have determined the transmission numerically using the standard non-equilibrium Green's function method \cite{datta2005quantum}.
As mentioned above, the transmission at the band center will eventually decay with $N$ because of the localized nature of the edge solution. This is clearly seen for the $N$ odd case shown in Fig.~\ref{transmission}. However, in notable contrast to the decoherence rate discussed earlier, there is no difference in the transmission with respect to the location of the edge solution (left or right). The $N$ even case is more interesting, in that there is a significant difference between the case where $r>\rc$ (no edge states) and that where $r<\rc$ (2 edge states), as seen in Fig.~\ref{transmission}. In the former case (no edge states), the transmission simply decays exponentially with $N$ due to the gap at zero energy. This is similar to the decoherence rate shown in Fig.~\ref{fig-decoherence}. In the latter case (two edge states), the transmission first increases with $N$, because the edge state energies converge to zero, before the transmission decays again (for $N\gtrsim 70$ in Fig.~\ref{transmission}) due to the localized nature of the edge states. It is worthwhile noting that the transmission in this case (2 edge states) is still five orders of magnitude larger than in the odd case, which only has one edge state. This is a consequence of the symmetry of each of the 2 edge states in the even case, which live at both edges (see figs. \ref{sketcha} and \ref{fig-allstates}.(b)). Indeed, the edge state solutions are either antisymmetric ($E>0$) or symmetric ($E<0$) in site position,  while the edge state in the odd case is restricted to either the left or right side of the SSH chain. The takeaway message here is that with a simple transmission probe it is not possible to distinguish the left from the right edge state, in stark contrast to our decoherence probe. 

In variants of the SSH model, it is also possible for the localized zero-energy mode to exist at a local topological defect (or soliton) when breaking the translational symmetry of the dimers. For example, instead of alternating $t_1$ and $t_2$ throughout the chain there could be two consecutive $t_1$'s before the alternation continues. This would lead to a similar exponentially localized mode at the defect. If the qubit could be scanned over such a chain, one could identify the position and properties of the topological defect. This assumes a small local coupling between the qubit an the SSH chain, as would be the case in a typical scanning probe experiment. Hence, using a qubit as a probe for topological edge states would be very interesting and could be extended to other toplogical excitations such as Majorana fermions with implications in quantum computing \cite{freedman2003topological}. 

Summarizing, we have shown that a qubit can be used as sensitive local detector of topological edge states by looking at its dynamics, without affecting the topological nature of the states. This has important implications on the experimental detection of topological states as well as, more generally, for the implementation of topological quantum computation.

\section*{Acknowledgments}
This work was supported in part by the Natural Science and Engineering Research Council of Canada and by the Fonds de Recherche Nature et Technologies du Qu{\'e}bec via the INTRIQ strategic cluster grant. RM is grateful for the hospitality of Perimeter Institute where part of this work was carried out. Research at Perimeter Institute is supported by the Government of Canada through the Department of Innovation, Science and Economic Development and by the Province of Ontario through the Ministry of Research, Innovation and Science.

\bibliographystyle{apsrev4-1}
\bibliography{refsSSH}

\end{document}